# Spin current relaxation time in thermally evaporated naphthyl diamine derivative films

Eiji Shikoh [a,b,*], Yuichiro Onishi [b], Yoshio Teki [a,c]

[a] *Graduate School of Engineering, Osaka Metropolitan University, Osaka, 558-8585, Japan*

[b] *Graduate School of Engineering, Osaka City University, Osaka, 558-8585, Japan*

[c] *Graduate School of Science, Osaka City University, Osaka, 558-8585, Japan*

[*]Corresponding author.

E-mail address: shikoh@omu.ac.jp (Eiji Shikoh)

*Abstract:*

The spin relaxation time ($\tau$) on the spin transport in thermally evaporated thin films of a naphthyl diamine derivative: *N, N'*-Bis(naphthalen-1-yl)-*N, N'*-bis(phenyl)-2,2'-dimethylbenzidine ($\alpha$-NPD) was evaluated with the spin-pump-induced spin transport properties and the electrical current-voltage properties in $\alpha$-NPD films. The zero-bias mobility and the diffusion constant of charges in $\alpha$-NPD films were obtained to be $(1.16\pm0.40)\times10^{-3}$ cm$^2$/Vs and $(2.97\pm1.02)\times10^{-5}$ cm$^2$/s, respectively. Using these values and the previously evaluated spin diffusion length in $\alpha$-NPD films of $62\pm18$ nm, the $\tau$ in $\alpha$-NPD films was estimated to be $1.90\pm1.41$ $\mu$s at room temperature, under an assumption of diffusive transport of the spin current in $\alpha$-NPD films. This estimated $\tau$ in $\alpha$-NPD films is long enough for practical use as a spintronic molecular material.

# 1. Introduction

Since organic molecular materials are soft, flexible, and environmental-friendly, those have been used as key-materials for next generation electronic devices [1]. Recently, to use the spin degrees of freedom in organic molecular materials are also focused, that is, not only molecular electronics but also molecular spintronics have been attracted much attention [2]-[9]. Organic molecular materials composed of light elements are promising candidates as spin transport materials because of their weak spin-orbit interaction working as spin scattering centers [2]-[9]. Therefore, to discover the molecular materials possessing longer spin diffusion length and/or longer spin relaxation time is one significant issue in the molecular spintronics.

For evaluation of the spin transport properties in organic molecular materials with low conductive materials, to use the spin-polarized current as spin injection method is generally hard because of the electrical conductance mismatch problem between the high conductive ferromagnetic metal as a spin injector and the low conductive material, which causes lowering the spin injection efficiency due to the spin scattering at the interface between those materials [10]-[11]. To suppress this electrical conductance mismatch problem in molecular spintronics, it is tried to make the hybridization of electronic orbit at the interface between the ferromagnetic metal as a spin injector and an organic molecular material by appropriate material choice for the hybridization, which is a so-called spinterface [2], [9], or to use tunnelling process with a tunnelling barrier at the interface between the ferromagnetic metal and an organic molecular material [2], [9], which is also a kind of spinterfaces. As another spin injection method to avoid the above electrical conductance mismatch problem at the interface, the spin-pump-induced spin current driven by the ferromagnetic resonance (FMR) [12] is focused since the electrical conductance mismatch problem in the spin injection process by the spin-pumping can be almost ignored [13]-[15]. Thus, the combination method of the spin pumping driven by FMR and the inverse spin-Hall effect (ISHE) [16] as a spin current detection method is widely used for evaluation of the spin transport properties in materials [14]-[15], [17]-[27]. At present, using the above combination method, the spin transport properties of some typical organic molecular films have been evaluated even at room temperature (RT) [15], [18]-[27].

Meanwhile, there is a simple question what carries spins (spin angular momenta) in the almost insulating materials like general organic molecular materials. Currently, two models as the spin transport mechanisms in organic molecular materials are considered: one is due to the polarons in the materials related to their low carrier densities, another is due to the exchange coupling between localized spins in the materials [15], [19], [28]. The pure spin current, which is the only flow of the spin angular momenta in a material, is generated under the spin-pumping regime, and this means that no bias voltages or electrical charge currents are applied in the material for the spin transport experiments. Thus, in this case, very few numbers of carriers in molecular materials diffusively carry their spins with forming polarons if the former model is dominant [15], [19], and almost carriers in organic molecular materials might be impurities even if organic molecular materials with high purity such as sublimed grade products are tested. Moreover, there is a possibility that the spin polarized carriers are directly injected from the ferromagnetic metal as the spin injector into the molecular materials in the spin pumping process [27]. Meanwhile, the latter model derived from the exchange coupling between localized spins would be appeared in the case that the carrier density is relatively large and then the average distance between localized spins might become relatively close [15], [19]. While localized spins in general organics are weak compared to inorganics, the exchange coupling between localized spins in organics wouldn't be ignored. At present, nobody distinguished which model is dominant as the spin transport mechanism in molecular materials. To discuss the spin transport mechanism in molecular materials more deeply, the experimental data to confirm are still lacked. Thus, in order to clarify the spin transport mechanism in molecular materials, the spin transport properties of more various molecular materials films, especially for pure spin current regime, must be experimentally investigated.

Recently, as one of typical molecular films, a thermally-evaporated amorphous thin film of a naphthyl diamine derivative: *N, N'*-Bis(naphthalen-1-yl)-*N, N'*-bis(phenyl)-2,2'-dimethylbenzidine (so-called as α-NPD), which is known as a typical hole transporting material in organic light emitting diodes (OLEDs: [29]-[30]) has been focused and the spin transport in α-NPD films has been demonstrated at RT with the combination method composed of the spin-pumping and the ISHE [27]. The spin diffusion length ($\lambda$) in α-NPD films was estimated to be 62±18 nm at RT by using the film thickness dependence of the spin transporting signals [27]. In this study, the spin relaxation time ($\tau$) on the spin transport in α-NPD films, that is, the spin current relaxation time in α-NPD films is evaluated with the spin-pump-induced spin transport properties and the electrical current-voltage properties in α-NPD films. The $\tau$ estimated in this study would correspond to a so-called transverse spin relaxation time in α-NPD films due to spin-lattice scattering on the spin transport.

## 2. Experimental methods

In this study, the $\tau$ in α-NPD films is estimated with the following procedure: Under an assumption of diffusive transport of the spin current by few numbers of carriers in α-NPD films [27], the relationship: [15], [22]

$$\tau = \frac{\lambda^2}{D}, \qquad (1)$$

in thermally evaporated α-NPD films is established, where $D$ is the diffusion constant of charges

(mainly, polarons) in the α-NPD films. Thus, the $\lambda$ and $D$ of α-NPD films are necessary to estimate the $\tau$ in α-NPD films. Figure 1 (a) shows a schematic illustration of our sample structure and experimental setup to previously demonstrate the spin transport in an α-NPD film [27]. Fig. 1(b) shows the sample dimensions on the top view. An external static magnetic field ($H$) is applied with an angle ($\theta$) to the sample film plane. The spin transport in an α-NPD film has been observed as follows: in a palladium (Pd)/α-NPD/$Ni_{80}Fe_{20}$ tri-layer structure sample, a pure spin current driven by the spin pumping with the FMR of the ferromagnetic $Ni_{80}Fe_{20}$ film is generated in the α-NPD layer. This spin current is then absorbed into the Pd layer. The absorbed spin current is converted into a charge current due to the ISHE in Pd and detected as an electromotive force (EMF). That is, to detect an EMF due to the ISHE in Pd under the FMR of the $Ni_{80}Fe_{20}$ film of the tri-layer structure samples is clear evidence for the spin transport in an α-NPD film. As mentioned before, the spin transport in α-NPD films has already been achieved with the combination method of the spin pumping and the ISHE, and the $\lambda$ in α-NPD films of 62±18 nm at RT has been evaluated [27]. Therefore, in this study, the obtained $\lambda$ with its deviation was used to estimate the $\tau$ in α-NPD films.

To obtain the $D$ in molecular films, the Einstein relationship: [31]

$$\frac{D}{\mu} = \frac{k_B T}{q}, \qquad (2)$$

is used which is satisfied even in disordered molecular films, where $\mu$, $k_B$, $T$ and $q$ are the zero-bias charge mobility (not a field effect mobility) in molecular films, the Boltzmann constant, the environmental temperature, and the elemental charge, respectively. The $\mu$ is obtained from measurements of the charge current density ($J$) versus the applied bias voltage ($V$) properties of the target molecular film. In an organic molecular film, the $J$ is approximately proportional to $V^2$ at lower bias region, which is called as the space charge limited current regime and is described as follows: [32]

$$J = \frac{9}{8}\varepsilon\mu\frac{V^2}{L^3}, \qquad (3)$$

where $\varepsilon$ is the dielectric constant in the molecular film and $L$ is the charge transport distance in the molecular film which corresponds to the $\alpha$-NPD film thickness of the tri-layer samples in our study. Using the eq.3, the $\mu$ in $\alpha$-NPD films is estimated. In the pure spin current regime, any electrical voltage or charge current is not applied in the pathway of the spin transport, as mentioned before. Therefore, the "almost zero-" bias charge mobility has been used to obtain the $D$, similarly to the previous studies [15], [19]-[23], [25].

Pd/$\alpha$-NPD/Ni$_{80}$Fe$_{20}$ tri-layer stacking samples as shown in Fig. 1 were formed as follows: A thermally oxidized silicon substrate was ultrasonically cleaned in isopropyl alcohol and then the substrate was dried out. Electron beam (EB) deposition was used to deposit Pd (Furuuchi Chemical Co., Ltd., 99.99% purity) to a thickness of 10 nm on the cleaned substrate, under a vacuum pressure of <10$^{-6}$ Pa. The deposition rate of Pd was controlled to be 0.05 nm/s with monitoring by a quartz resonator while the substrate temperature was not controlled during Pd depositions. Next, also under a vacuum pressure of <10$^{-6}$ Pa, $\alpha$-NPD molecules (Tokyo Chemical Industry Co., Ltd.; sublimed grade) were thermally evaporated to a thickness of $d$ ($d = 25\sim100$ nm) through a shadow mask. The deposition rate of $\alpha$-NPD molecules was controlled to be 0.1 nm/s with monitoring by a quartz resonator while the substrate temperature was not controlled during $\alpha$-NPD depositions. Finally, Ni$_{80}$Fe$_{20}$ (Kojundo Chemical Lab. Co., Ltd., 99.99%) was deposited to a thickness of 25 nm by EB deposition through another shadow mask, under a vacuum pressure of <10$^{-6}$ Pa. During Ni$_{80}$Fe$_{20}$ depositions, the deposition rate was controlled to be 0.03 nm/s with monitoring by a quartz resonator, and the sample substrate was controlled with a cooling medium of -2°C to prevent the deposited molecular films from breaking.

The $J$-$V$ properties in an $\alpha$-NPD film were measured by using the same stacking structure samples with the $d$ of 50 nm, as shown in Fig. 1 (a). For the measurement of $J$-$V$ properties of an α-NPD film, a two-terminal method was used with a DC voltage current source/monitor (ADVANTEST Corporation, R6243). One leading wire was attached at the top ($Ni_{80}Fe_{20}$) layer and another wire was attached at the bottom (Pd) layer of the stacking structure samples, as shown in Fig. 1(c). All measurements were performed at RT.

## 3. Results and discussion

Figure 2 shows (a) a typical $J$-$V$ property and (b) the $J$-$V^2$ property of an α-NPD film of a stacking structure sample. The dashed line in Fig. 2(b) shows a linear fitting to the $J$-$V^2$ properties at lower bias, which has been drawn to the range being proportional to $V^2$ on the experimental data. Using the eq. (3) with the relative dielectric constant ($\varepsilon_r$) in α-NPD films of ~2.3 [33], the $\mu$ in α-NPD films was estimated to be $(1.16\pm0.40)\times10^{-3}$ $cm^2$/Vs at RT. Here, just for simplification, the difference between the work function of the metals ($Ni_{80}Fe_{20}$ or Pd) and the ionization potential of the α-NPD film is ignored because only the divergence of $J$-$V^2$ properties at low bias is used for the analysis. The deviation on the $\mu$ estimation comes from the deviation of the divergence of the linear fitting to the $J$-$V^2$ properties. Under an assumption of diffusive transport of the spin current in α-NPD films, the $D$ in α-NPD films has been calculated to be $(2.97\pm1.02)\times10^{-5}$ $cm^2$/s with the eq. (2). The deviation on the $D$ calculations is due to the deviation of the $\mu$ estimations. Eventually, using the eq. (1) and the previously obtained $\lambda$ of $62\pm18$ nm at RT in thermally evaporated α-NPD amorphous films [27], the $\tau$ in α-NPD films was estimated to be $1.90\pm1.41$ μs at RT. In the deviation of this estimated $\tau$, the deviation of the previously estimated $\lambda$ in α-NPD films is also included, that is, the possible spin scattering origins on the spin transport in α-NPD films (the scatterings due to vacancies, impurities, grain boundaries in the amorphous α-NPD films, and so on), are considered in this $\tau$ estimation.

Here, the validity of the $\tau$ estimation in α-NPD films is discussed. Because the spin relaxation time in α-NPD films evaluated with other experimental methods has been absent yet, the estimated $\tau$ of α-NPD films in this study is compared with the $\tau$ of other molecular films estimated with the same experimental method, that is, the spin-pump-induced spin injection experiments. Figure 3 shows a relationship between the reported $\tau$ and the $\lambda$ of some typical molecular films (PEDOT:PSS [34], doped PEDOT:PSS [20], PBTTT [15], doped PBTTT [23], Alq$_3$ [19], pentacene [21]-[22], and rubrene [25]), including α-NPD films in our study. Thinking from the suggested spin transport mechanism in organic molecular materials, not only the estimated long $\tau$, but also the previously estimated long $\lambda$ for α-NPD films may be surprising because, first, no electrical bias has been applied in the spin-pump-induced spin injection, which is different from, for instance, OLED operations. Second, α-NPD films are amorphous [27], [35], where the electronic orbit make random networks. The boundaries between molecules may dominantly work as spin scattering centers in spin transport in such amorphous films. However, for instance, the Alq$_3$ films are also amorphous which has longer $\tau$ than α-NPD and almost comparable $\lambda$ to α-NPD [19]. Thus, it is concluded the estimated $\tau$ in α-NPD films is valid.

From Fig. 3, some interesting trend can be considered. Except for some materials (the open squares), the $\tau$ approximately tends to be proportional to the $\lambda^2$, which corresponds to the eq (1). In this point, the $\tau$ estimation for α-NPD films would be valid. However, this is natural because the diffusive spin transport in $\tau$ estimations for all materials in Fig. 3 has been assumed. Second, polymer films [15], [20], [23], [34] tend to possess longer spin diffusion length than small molecular films [19], [21]-[22], [25]. The difference between polymer and small molecular films would be explained with the difference of the grain boundary amount working at spin scattering centers, that is, small molecular films have more grain boundary than polymer films. This point can be confirmed by using single-crystal-like molecular films, if available. Next, the molecular films with higher carrier density (carrier-doped films corresponding to the open squares in Fig. 3) tend to possess shorter $\tau$ than non-doped films (solid squares). This reason might be due to the increase of carriers working as spin scattering centers with the carrier doping, and the spin transport properties of the molecular films with higher carrier density might be apart from the relationship of the eq (1). In the carrier-doped cases, the suggested spin transport mechanism due to the exchange coupling between spins might be strongly affected. PEDOT:PSS films with higher carrier density possess longer $\lambda$ than non-doped PEDOT:PSS films, while the $\lambda$ of PBTTT films looks almost independent of the carrier density. In this case, too, the exchange coupling induced spin transport might be related. Although the polymer data in Fig. 3 are obtained by other research groups, those results depend on materials, in short, the dominant spin transport mechanism might depend on materials, especially on the carrier densities. In other words, by changing the carrier density in almost insulating molecular materials with an external field (for instance, electrical field, heat, light irradiation, and so on), there is possibility that the spin transport properties in molecular films can be controlled with an external field, which is in spite of the molecular weight. While the carrier density control in molecular materials with an external field is out of this study, it will probably be confirmed near future. The estimated $\tau$ of 1.90±1.41 μs and $\lambda$ of 62±18 nm in thermally evaporated α-NPD films in this paper are moderate among the reported values of the molecular films, but, long enough for practical use as a spin transport material for spintronic devices among today's industrial technology. The spin transport mechanism in molecular films is still unclear and should be kept studying, whereas the spin-pump-induced spin transport properties in α-NPD films have been clarified, that is, one of the expected data in the molecular spintronics is successfully obtained.

**4. Conclusions**

The spin current relaxation time in thermally evaporated thin α-NPD films was evaluated with the spin-pump-induced spin transport properties and the current-voltage properties in α-NPD films. The zero-bias mobility and the diffusion constant of charges in α-NPD films were experimentally obtained to be $(1.16\pm0.40)\times10^{-3}$ cm$^2$/Vs and $(2.97\pm1.02)\times10^{-5}$ cm$^2$/s, respectively. Using those values and the previously obtained spin diffusion length in α-NPD films of $62\pm18$ nm, the spin current relaxation time in α-NPD films was estimated to be $1.90\pm1.41$ μs at room temperature, under an assumption of a diffusive spin transport in α-NPD films. This estimated spin current relaxation time in α-NPD films is long enough for practical use as a spin transport material for spintronic devices.

**Author statement**

Eiji Shikoh: Supervising, Data analysis, Discussion, Manuscript Writing

Yuichiro Onishi: Experiments, Data analysis, Discussion

Yoshio Teki: Discussion

**Declaration of competing interest**

The authors declare no competing financial interests.

**Acknowledgments**

This research was partly supported by the Grant-in-Aid from the Japan Society for the Promotion of Science (JSPS) for Scientific Research (B) (No. 20H02715) (to Y. T and E. S.), by the Grant-in-Aid from the JSPS for Scientific Research (C) (No. 23K04569) (to E. S.), and by the Cooperative Research Program of "Network Joint Research Center for Materials and Devices"

(to E. S.).

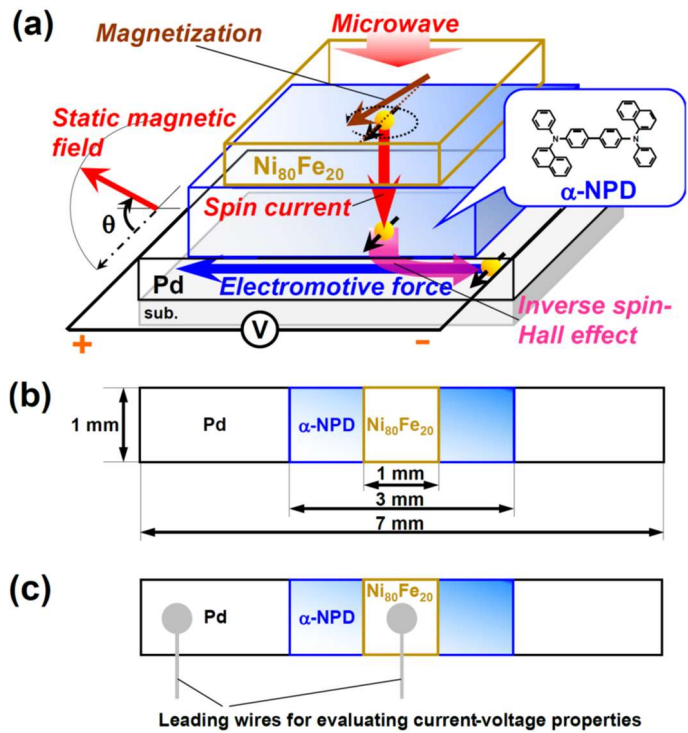

Fig. 1. (a) A schematic illustration of our sample structure and experimental setup for the spin transport demonstration. (b) Dimensions with a top view of our samples. (c) A schematic illustration of leading wire setting to our sample for evaluation of current-voltage properties with a two-terminal method.

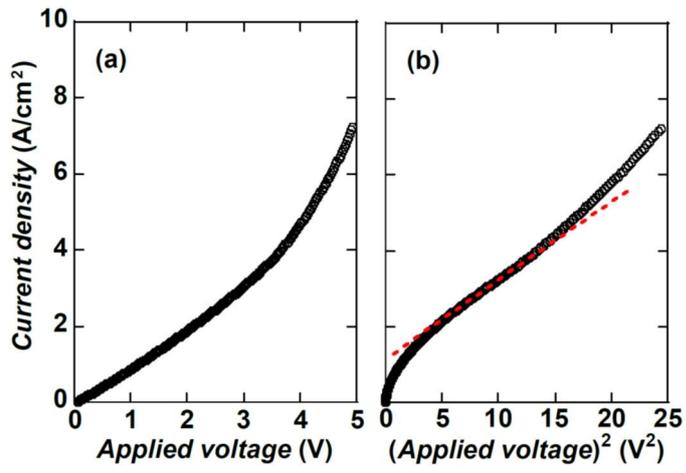

Fig. 2. (a) A current density ($J$) vs Applied bias voltage ($V$) property and (b) the $J$-$V^2$ property of an a-NPD film. The dashed line in (b) shows a linear fitting to the $J$-$V^2$ properties at low bias, drawn to the proportional range to $V^2$.

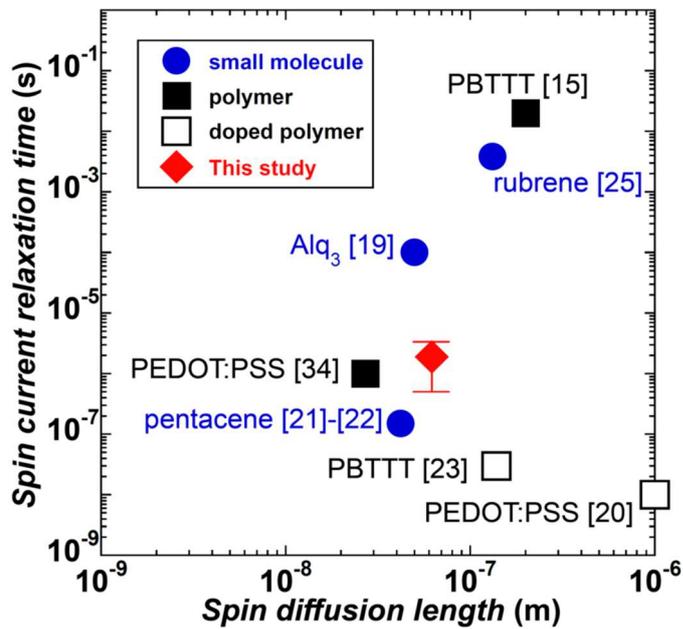

Fig. 3. A relationship between the reported spin relaxation time and the spin diffusion length of molecular films evaluated with the spin pumping experiments. The numbers in parentheses are reference data.